\documentclass{jnmp}

\usepackage{amsmath}
\JNMPnumberwithin{equation}{section}

\theoremstyle{definition}

\newcommand{\cS}{{\mathcal S}}
\def\ket#1{|#1\rangle}
\newcommand{\fB}{{\mathfrak B}}

\newcommand{\tS}{\tilde{S}}

\newcommand{\La}{\Lambda}
\newcommand\hLa{\widehat\La}
\DeclareMathOperator\sgn{sgn}
\newcommand{\ep}{\epsilon}
\newcommand{\pa}{\partial}
\newcommand{\cH}{{\mathcal H}}
\newcommand{\tC}{\widetilde{C}}
\newcommand{\bx}{\mathbf{x}}

\newcommand\ZZ{\mathbb Z}
\newcommand\RR{\mathbb R}
\newcommand{\tK}{\tilde{K}}
\newcommand{\cR}{{\mathcal R}}
\newcommand{\Si}{\Sigma}
\newcommand{\card}{\operatorname{card}}
\newcommand\e{\mathrm e}
\newcommand\I{\mathrm i}
\newcommand{\cB}{{\mathcal B}}
\newcommand\AND{\;\mathrm{and}\;}

\newcommand{\de}{\delta}
\newcommand{\la}{\lambda}

\newcommand{\si}{\sigma}
\newcommand\ssh{\mathsf{h}}
\newcommand{\ssZ}{\mathsf{Z}}
\newcommand\Hs{H_\text{s}}
\newcommand\Zs{Z_\text{s}}
\newcommand{\Bbe}{\overline\beta}
\newcommand\cP{\widetilde{\mathcal P}}

\newcommand\NN{\mathbb N}

\newcommand{\bfB}{\boldsymbol\fB}
\newcommand{\bSi}{\boldsymbol\Sigma}
\newcommand{\ze}{\zeta}
\newcommand{\ms}{\mspace{1mu}}
\def\Hs{H_{\mathrm{s}}}

\begin{document}

%
\renewcommand{\evenhead}{A. Enciso, F. Finkel, A. Gonz{\'a}lez-L\'opez, and M.A.
  Rodr{\'\i}guez} \renewcommand{\oddhead}{A Haldane--Shastry spin chain of $BC_N$
  type in a constant magnetic field}

%
\thispagestyle{empty}

\FirstPageHead{*}{*}{20**}{\pageref{firstpage}--\pageref{lastpage}}{Article}

\copyrightnote{2004}{A. Enciso, F. Finkel, A. Gonz{\'a}lez-L{\'o}pez and M.A. Rodr{\'\i}guez}

\Name{A Haldane--Shastry spin chain of $BC_N$ type in a constant magnetic field}

\label{firstpage}

\Author{A. ENCISO, F. FINKEL, A. GONZ{\'A}LEZ-L\'OPEZ~$^\dag$, and M.A. RODR{\'I}GUEZ}

\Address{Departamento de F{\'\i}sica Te\'orica II, Universidad Complutense, 28040 Madrid, Spain \\
$^\dag$E-mail: artemio@fis.ucm.es\\[10pt]}

\Date{Received Month *, 200*; Revised Month *, 200*;
Accepted Month *, 200*}

\begin{abstract}
\noindent We compute the spectrum of the trigonometric Sutherland
spin model of $BC_N$ type in the presence of a constant magnetic
field. Using Polychronakos's freezing trick, we derive an exact
formula for the partition function of its associated
Haldane--Shastry spin chain.
\end{abstract}

%
\section{Introduction}

In 1971, F. Calogero \cite{Ca71} introduced a solvable quantum model
describing a system of $N$ particles with two-body interactions depending on
the inverse square of the particles' distance. In the same year, B. Sutherland
\cite{Su71} proposed a similar model with an interaction potential of
trigonometric type. The importance of these models, which was already apparent
from the very beginning, is now widely acknowledged by the theoretical and
mathematical physics community. {}From a mathematical point of view, these
models are integrable both in their classical and quantum versions, in the
sense that they admit a complete set of integrals of motion. Moreover, they
are also exactly solvable, in the sense that their spectrum and eigenfunctions
can be expressed in closed form.  {}From a more physical point of view,
Calogero--Sutherland (CS) models play an important role in many different
fields, like for instance Yang--Mills theories~\cite{GN94}, the quantum Hall
effect~\cite{AI94}, random matrices~\cite{SLA94,TSA95} and fractional
statistics~\cite{Ha94,Po89}.

Calogero--Sutherland models were cast into a very
elegant mathematical framework by Olshanetski and Perelomov in
\cite{OP83}. They showed that these models are limiting cases of a
more general one with a two-body interaction potential of elliptic
type, and uncovered their relation to the $A_N$ root system. In
fact, these authors also constructed generalizations of the
previous models associated with all the classical (extended) root
systems, like $BC_N$.

The extension of CS models to particles with internal degrees
(typically interpreted as spin) of freedom was first proposed in
the classical case in Ref.~\cite{GH84}. In the quantum case, spin
CS models were actively studied in the last decade, both in the
$A_N$ and $BC_N$ cases. It turns out that these models inherit the
basic properties of their scalar counterparts, namely their
integrability and exact-solvability. There are essentially two
approaches in the study of spin CS models, namely the
supersymmetric formalism~\cite{BTW98,DLM01,Gh01} and the Dunkl or
exchange operator
method~\cite{Ba96,BGHP93,Du98,FGGRZ01,FGGRZ01b,Po92,Ya95}.
Usually, but not always~\cite{Po94}, the interaction of the spins
with a constant external magnetic field is not included in the
spin CS Hamiltonian.

In 1988, Haldane \cite{Ha88} and Shastry \cite{Sh88} independently introduced
a new type of solvable spin chain with long-range position-dependent
interactions. The sites of this chain are equidistant points in a circle, and
the interaction between the corresponding spins is proportional to the inverse
square of their chord distance. Shortly afterwards, Fowler and Minahan
\cite{FM93} showed that this chain is completely integrable by means of
Polychronakos's exchange operator formalism \cite{Po92}. The connection with
the Sutherland model, although already noted by Shastry in his paper, was made
precise by Polychronakos in \cite{Po93}, using what he called the ``freezing
trick". The main idea behind this method, which can actually be applied to any
spin CS model, is to take the strong coupling constant limit in the
Hamiltonian, so that the particles become ``frozen" at the equilibrium
positions of the scalar part of the potential. In this way one can obtain new
spin chains of Haldane--Shastry (HS) type, in which the sites are not
necessarily equally spaced. Most of the literature on HS spin chains is
devoted to those based on dynamical spin models of $A_N$ type, while their
$BC_N$ counterparts have received comparatively less attention. Yamamoto and
Tsuchiya proved the integrability of the rational HS chain of $BC_N$ type
\cite{YT96}, although they did not compute its spectrum. The trigonometric
$BC_N$ spin chain was discussed by Bernard, Pasquier and Serban \cite{BPS95}
in the spin $1/2$ ferromagnetic case, but only for equally spaced sites. The
integrability of the trigonometric/hyperbolic version of this chain was
established in \cite{CC04,FGGRZ03}, although again its spectrum was not
computed.  In a recent work \cite{EFGR04}, we have extensively studied the
trigonometric $BC_N$ spin chains (both ferromagnetic and antiferromagnetic)
for arbitrary spin and without assuming that the sites are equally spaced. The
main result in the latter paper is the derivation of a closed-form expression
for the partition function of the model, following a method based on
Polychronakos's freezing trick.

In this paper we study the effect of the presence of a constant
external magnetic field in the trigonometric $BC_N$ Sutherland
model for spin $1/2$ and in its associated spin chain. This
modifies the Hamiltonian of the spin chain by the addition of a
term proportional to the projection of the total spin operator
along the direction of the magnetic field. It turns out that both
the dynamical spin model (if the magnetic field is suitably
oriented) and its associated chain remain solvable, although the
method used in \cite{EFGR04} to compute the spectrum of the
dynamical model, based on expressing the Hamiltonian in terms of a
commuting family of Dunkl operators, cannot be applied in this
case. We shall see, however, that the basis of the Hilbert space
constructed in the latter reference to solve the model in the
absence of a magnetic field can be slightly modified so that the
Hamiltonian of the dynamical model is still triangular. {}From the
spectrum of the dynamical model we shall then evaluate in closed
form the partition function of its associated spin chain using the
freezing trick.

The paper is organized as follows. In Section 2 we introduce the
dynamical spin model and its corresponding Hilbert space. The
spectrum of this model is computed in Section~3, by showing that
the term due to the magnetic field is triangular in an appropriate
modification of the basis constructed in~\cite{EFGR04}. In
Section~4 we define the HS spin chain and compute its partition
function using the freezing trick and the results of the previous
section.

\section{Preliminary definitions}

In this section we set up the notation used throughout the paper
and define the Hamiltonian of the Sutherland spin model of $BC_N$
type in the presence of a constant external magnetic field. We
shall denote by $\cS$ the Hilbert space corresponding to the
internal degrees of freedom of $N$ identical spin $\frac12$
particles. Let
\begin{equation}\label{basis}
  \cB_\cS^\ze=\big\{ \ket
  s\equiv\ket{s_1,\dots,s_N}\;\big|\;s_i={}\uparrow\downarrow\!\big\}\,,
\end{equation}
be a basis of $\cS$ whose elements are simultaneous eigenstates of the $\ze$
component of the one-particle spin operators, $O\ze$ being an arbitrary
direction. The corresponding spin permutation and reversal operators $S_{ij}$
and $S_i$ are defined by
\begin{equation}
\begin{aligned}
  & S_{ij}|s_1,\dots,s_i,\dots,s_j,\dots,s_N\rangle=|s_1,\dots,
  s_j,\dots,s_i,\dots,s_N\rangle\,,\\
  &
  S_i|s_1,\dots,s_i,\dots,s_N\rangle=|s_1,\dots,-s_i,\dots,s_N\rangle\,.
\label{SS}
\end{aligned}
\end{equation}
We will also use the customary notation $\tS_{ij}=S_iS_jS_{ij}$. The operators
$S_i$ and $S_{ij}$ generate a multiplicative group isomorphic to the
Weyl group of $B_N$ type. Similarly, the coordinate permutation and
sign-reversal operators $K_{ij}$ and $K_i$ are defined by
\begin{align*}
&(K_{ij}f)(x_1,\dots,x_i,\dots,x_j,\dots,x_N)=f(x_1,\dots,
x_j,\dots,x_i,\dots,x_N)\,,\\
&(K_i f)(x_1,\dots,x_i,\dots,x_N)=f(x_1,\dots,-x_i,\dots,x_N)\,,
\end{align*}
and $\tK_{ij}=K_iK_jK_{ij}$. The total permutation and sign reversal operators
will be denoted by $\Pi_i\equiv K_iS_i$ and $\Pi_{ij}\equiv K_{ij}S_{ij}$. The
multiplicative group generated by $K_i$ and $K_{ij}$ (respectively, $\Pi_i$ and
$\Pi_{ij}$) is also isomorphic to the $B_N$-type Weyl group.

Let us define the antisymmetrizer with respect to the symmetric
group generated by $\Pi_{ij}$ as
\[
\hLa=\frac1{N!}\sum_{i=1}^{N!}\sgn(P_i)\,P_i\,,
\]
$P_i$ being an element of this group and $\sgn(P_i)$ its
signature. Likewise, we will denote by
\[
\hLa_\ep=2^{-N}\prod_i(1+\ep\,\Pi_i)
\]
the symmetrization ($\ep=1$) or antisymmetrization ($\ep=-1$) with respect to
sign reversals. Here and in what follows all sums and products run from 1 to
$N$, unless otherwise stated. We shall make use of the projection operator
\[
\La_\ep=\hLa\,\hLa_\ep
\]
on states antisymmetric under particle permutations and with
parity $\ep$ under sign-reversals. The Hamiltonian of the spin
Sutherland model of $BC_N$ type in a constant external magnetic
field $\bfB$ is given by
\begin{equation}
\label{H}
\begin{aligned}
  H_\ep=&-\sum_i \pa_{x_i}^2 + a\,\sum_{i\neq j}\left[\sin^{-2}
    (x_i-x_j)\,(a+S_{ij})+\sin^{-2} (x_i+x_j)\,(a+\tS_{ij})\right]\\
  &{}+b\,\sum_i \sin^{-2}\!x_i\,(b-\ep S_i)+b'\,\sum_i
  \cos^{-2}\!x_i\,\big(b'-\ep S_i\big)\\
  &{}-eg\,\bfB\cdot\bSi+\frac{e^2}4(\fB_y^2+\fB_z^2)\sum_ix_i^2\,,
\end{aligned}
\end{equation}
where $\bSi=(\Si^\xi,\Si^\eta,\Si^\ze)$ is the total spin operator and ($O\xi,
O\eta,O\ze)$ is an arbitrary system of orthogonal axes. Here we have assumed
that the real constants $a,b,b'$ are greater than $\frac12$, and we have
denoted by $g$ and $e$ the particles' gyromagnetic ratio and electric charge,
respectively. The last (diamagnetic) term in Eq.~\eqref{H} has to be dropped
to preserve the solvability of $H_\ep$. In fact, this term vanishes
identically if the magnetic field is parallel to the $x$ axis. Since our main
interest is to study the spin chain associated with the Hamiltonian $H_\ep$,
we have preferred in what follows to drop the diamagnetic term in order to
keep the direction of the magnetic field in the spin chain arbitrary.

The above Hamiltonian possesses inverse-square type
singularities for $x_i\pm x_j=k\pi$ and $x_i=k\pi/2$, with $k\in\ZZ$. Since the
nature of these singularities prevents the particles from overtaking each
other and from crossing the singular hyperplanes $x_i=k\pi/2$, the particles
may be regarded as distinguishable, with configuration space
\begin{equation}\label{tC}
\tC=\Big\{\bx=(x_1,\ldots,x_N)\in \RR^N\;\Big|\;0<x_1<\cdots
<x_N<\frac\pi2\Big\}\,.
\end{equation}
The Hilbert space of the system can thus be taken as
$\cH=L^2_0(\tC)\otimes\cS$, with
\begin{multline*}
  L^2_0(\tC)=\Big\{f\in L^2(\tC)\;\Big|\;\exists\varlimsup_{x_i\pm x_j\to k\pi}
  |x_i\pm x_j-k\pi|^{-a}|f(\bx)|\,,\quad\exists\varlimsup_{x_i\to 0} |x_i|^{-b}|f(\bx)|\,,\\
  \exists\varlimsup_{x_i\to\pi/2} |x_i-\pi/2|^{-b'}|f(\bx)|\,; \quad
  k=0,1,\quad 1\leq i\neq j\leq N\Big\}\,.
\end{multline*}
The conditions imposed in the definition of $L^2_0(\tC)$ guarantee
the finiteness of $(\psi,H_\ep\psi)$ for all $\psi\in\cH$. It can
be shown that the operator $H_\ep:\cH\to\cH$ is actually
equivalent to any of its extensions to spaces of symmetric or
antisymmetric functions (with respect to both permutations and
sign reversals) in $L_0^2(C)\otimes\cS$, where $C$ is the $N$-cube
$(-\frac\pi2,\frac\pi2)^N$ and $L_0^2(C)$ is defined similarly to
$L^2_0(\tC)$. For technical reasons, it is convenient to consider
that $H_\ep$ acts in the Hilbert space
\begin{equation}
\cH_\ep=\La_\ep\big(L^2_0(C)\otimes\cS\big)
\end{equation}
of states antisymmetric under permutations and with parity $\ep$ under sign
reversals.

\section{Spectrum of the dynamical model}

The Hamiltonian~\eqref{H} with $\bfB=0$ was thoroughly studied in
\cite{EFGR04}, and its spectrum was exactly computed. The calculation was
based on the fact that when $\bfB=0$ the Hamiltonian \eqref{H} is the image of
the operator
\begin{equation}
\label{HpB}
\begin{aligned}
  H'_0=-\sum_i \pa_{x_i}^2 &+ a\,\sum_{i\neq j}\left[\sin^{-2}
    (x_i-x_j)\,(a-K_{ij})+\sin^{-2} (x_i+x_j)+\,(a-\tK_{ij})\right]\\
  &{}+b\,\sum_i \sin^{-2}\!x_i\,(b-K_i)+b'\,\sum_i
  \cos^{-2}\!x_i\,\big(b'-K_i\big)
\end{aligned}
\end{equation}
under the mapping (cf.~\cite{EFGR04})
\begin{equation}
  K_{ij}\mapsto-S_{ij},\qquad K_{i}\mapsto\ep S_{i}\,.\label{starmap}
\end{equation}
The operator $H'_0$ was expressed as a sum of squares of an
appropriate set of commuting Dunkl operators $J_i$, $i=1,\dots,N$,
which preserve a flag $\cR_0\subset\cR_1\subset\cdots$ of
finite-dimensional spaces of smooth functions, where
\begin{equation}\label{Rdef}
\cR_k = \big\langle\phi(\bx)\exp\big(2\I\sum_j n_j
x_j\big)\;\big|\; n_j=-k,-k+1,\dots, k\,,\quad j=1,\dots,N
\big\rangle\,,
\end{equation}
and
\begin{equation}\label{phi}
\phi(\bx) = \prod_{i<j}\big|\sin (x_i-x_j)\, \sin
(x_i+x_j)\big|^a\, \prod_i|\sin x_i|^b |\cos x_i|^{b'}\,.
\end{equation}
For each value of $k$, we constructed a basis $\cB_k$ of $\cR_k$
such that the matrix representing $H'_0$ in this basis is
triangular. Since the closure of
$\bigcup\limits_{k=0}^\infty\cR_k$ is $L^2_0(C)$, this observation
immediately yields the spectrum of $H'_0$. The spectrum of $H_\ep$
in the absence of magnetic field was then computed by suitably
extending the basis $\bigcup\limits_{k=0}^\infty\cB_k$ to a basis
$\cB^\ze_\ep$ of the Hilbert space $\cH_\ep$; see~\cite{EFGR04}
for the details.

The operators $S_{ij}$ and $S_i$ can be expressed in terms of
the usual one-particle spin operators $\bSi_i=(\Si^\xi_i,\Si^\eta_i,\Si^\ze_i)$
as follows
\[
S_{ij}=2\bSi_i\cdot\bSi_j+\frac12\,,\qquad S_i=2\ms\Si^\xi_i\,.
\]
Hence, if the magnetic field in the Hamiltonian~\eqref{H} is
directed along the $\xi$ axis, the term $\bfB\cdot\bSi$ can be
expressed in terms of the spin reversal operators $S_i$ as
$\frac12\fB\sum_iS_i$, where $\fB=\vert\bfB\vert$. In this case,
the Hamiltonian $H_\ep$ may be written as
\begin{align}
  H_\ep=&-\sum_i \pa_{x_i}^2 + a\,\sum_{i\neq j}\left[\sin^{-2}
    (x_i-x_j)\,(a+S_{ij})+\sin^{-2} (x_i+x_j)\,(a+\tS_{ij})\right]\notag\\
  &{}+b\,\sum_i \sin^{-2}\!x_i\,(b-\ep S_i)+b'\,\sum_i
  \cos^{-2}\!x_i\,\big(b'-\ep S_i\big)-\frac{eg}2\,\fB\sum_iS_i\,.\label{H2}
\end{align}
Although this Hamiltonian is the image of the operator
\begin{align*}
  H'_\ep=&-\sum_i \pa_{x_i}^2 + a\,\sum_{i\neq j}\left[\sin^{-2}
    (x_i-x_j)\,(a-K_{ij})+\sin^{-2} (x_i+x_j)\,(a-\tK_{ij})\right]\\
  &{}+b\,\sum_i \sin^{-2}\!x_i\,(b-K_i)+b'\,\sum_i
  \cos^{-2}\!x_i\,\big(b'-K_i\big)-\frac{\ep eg}2\,\fB\sum_iK_i\,,
\end{align*}
under the mapping \eqref{starmap}, the Dunkl operator techniques
of Ref.~\cite{EFGR04} cannot be directly applied to prove its
integrability, since it is not clear how to express the operator
$H'_\ep$ in terms of a commuting family of differential-difference
operators. Moreover, even if the new term proportional to the
magnetic field leaves invariant the spaces $\cR_k$, the matrix of
$H'_\ep$ in the basis $\bigcup\limits_{k=0}^\infty\cB_k$ of
$L^2_0(C)$ introduced in Ref.~\cite{EFGR04} need not be
triangular. It is more convenient, therefore, to work directly
with the spin Hamiltonian~\eqref{H2} and its representation in a
slight modification of the spin basis $\cB^\ze_\ep$ used in the
latter reference.

Let us recall, to begin with, the construction of the basis
$\cB^\ze_\ep$. To this end, we need to introduce the following
notation. Let $\NN_0^N$ denote the set of nonincreasing
multiindices $n=(n_1,\dots,n_N)$, with $n_i=0,1,\dots$ and
$n_1\geq\cdots\geq n_N$. If $n,n'\in\NN_0^N$, we shall say that
$n\prec n'$ if $n_1-n'_1=\cdots=n_{i-1}-n'_{i-1}=0$ and
$n_i<n'_i$. Let $f_n$ denote the function
\[
f_n(\bx)=\phi(\bx)\,\e^{2\I \sum_in_ix_i}\,,\qquad n\in\NN_0^N\,,
\]
where $\phi(\bx)$ is given by~\eqref{phi}. The basis $\cB^\ze_\ep$
can be taken as any linearly independent subset of the set
\begin{equation}\label{theset}
\big\{ \La_\ep(f_n \ket s)\;\big|\;n\in\NN_0^N\,,\: \ket s\in\cB^\ze_\cS\,\big\}
\end{equation}
ordered so that $\La_\ep(f_n \ket s)$ precedes
$\La_\ep(f_{n'} \ket{s'})$ if $n\prec n'$. It should be noted
that the basis $\cB^\ze_\cS$ in Eq.~\eqref{theset} could actually
be replaced by any basis of the spin space $\cS$ without changing
the triangularity of the Hamiltonian~$H_\ep\big|_{\fB=0}$.
As shown in Ref.~\cite{EFGR04},
a linearly independent subset of the set~\eqref{theset} is obtained
by imposing the following conditions on the multiindex $n$ and the
spin basis element $\ket s$:
\begin{itemize}
  \item[1.] $\#(m)\equiv\card\{i\;|\; n_i=m\}\leq2-\delta_{m0}$ for all
  $m=0,1,\dots$;
  \item[2.] If $n_i=n_{i+1}$, then $s_i={}\uparrow$ and $s_{i+1}={}\downarrow$;
  \item[3.] If $n_N=0$, then $s_N={}\uparrow$.
\end{itemize}

It is more convenient for our purposes to work a slight
modification $\cB^\xi_\ep$ of the basis $\cB^\ze_\ep$, obtained by
replacing $\cB_\cS^\ze$ in the previous construction by the basis
\begin{equation}\label{xbasis}
  \cB_\cS^\xi=\big\{\ms\ket
  \si\equiv\ket{\si_1,\dots,\si_N}\;\big|\;\si_i=\pm1/2\,\big\}
\end{equation}
of simultaneous eigenstates of the one-particle spin operators
$\Si^\xi_i$. Conditions 2 and 3 above should accordingly be replaced
by
\begin{itemize}
  \item[$2'$.] If $n_i=n_{i+1}$, then $\si_i=+1/2$ and $\si_{i+1}=-1/2$;
  \item[$3'$.] If $n_N=0$, then $\si_N=\ep/2$.
\end{itemize}
The last condition is due to the fact that the one-particle spin states
\[
\ket{\!\pm\!1/2\,}=\frac1{\sqrt2}\,\big(\,\ket{{}\uparrow{}}\pm\ket{{}\downarrow{}}\big)
\]
have parity $\ep$ under flipping of the $\ze$ component of the spin.
Since the operator $\sum_iS_i$ commutes with the
projector $\La_\ep$, it is diagonal in the basis $\cB^\xi_\ep$.
Indeed,
\[
\Big(\sum_iS_i\Big)\La_\ep\big(f_n \ket\si\big)=\La_\ep\sum_if_n\big(S_i\ket\si\big)
=\Big(\sum_i2\si_i\Big)\La_\ep\big(f_n \ket\si\big)\equiv\la(n,\si)\La_\ep\big(f_n \ket\si\big)\,.
\]
Taking into account the conditions $1$, $2'$, and $3'$ for the modified basis
$\cB^\xi_\ep$, it immediately follows that the eigenvalue $\la(n,\si)$ can be expressed as
\[
\la(n,\si)=d_+(n,\si)-d_-(n,\si)\,,
\]
where
\[
d_\pm(n,\si)=\card\big\{i\;\big|\; \#(n_i)=1\,\AND\,\si_i=\pm1/2\big\}\,.
\]
By the previous remark, the Hamiltonian $H_\ep\big|_{\fB=0}$ is still
triangular in the modified basis $\cB^\xi_\ep$, with diagonal elements
\begin{equation}\label{En0}
E_n^0=\sum_i\big(2n_i+b+b'+2a(N-i)\big)^2\,,
\end{equation}
cf. Ref~\cite{EFGR04}. It follows that the complete Hamiltonian~\eqref{H2}
is triangular in the basis $\cB^\xi_\ep$, with eigenvalues $E_{n\si}$ given by
\begin{equation}\label{Ens}
  E_{n\si}=E^0_n-\frac{eg}2\,\fB\,\la(n,\si)\,.
\end{equation}
This formula for the spectrum of the Hamiltonian~\eqref{H2} will be used in next section
to compute the partition functions of the corresponding spin chains.

\section{An HS spin chain of $BC_N$ type in a magnetic field}

Using Polychronakos's freezing trick \cite{Po93,Po94}, one may
obtain a Haldane--Shastry spin chain of $BC_N$ type associated
with the Hamiltonian~\eqref{H2}. This technique, thoroughly
discussed in \cite{EFGR04}, consists in taking the large coupling
constant limit $a\to+\infty$, while maintaining constant the ratios $\beta\equiv b/a$,
$\beta'\equiv b'/a$ and $B\equiv-eg\fB/(16a)$. In this limit, the eigenfunctions of
$H_\ep$ become sharply peaked around a minimum of the potential
\begin{equation}\label{U}
U(\bx)=\sum_{i\neq
  j}\big(\sin^{-2}(x_i-x_j)+\sin^{-2}(x_i+x_j)\big)
  +\sum_i(\beta^2\sin^{-2}x_i+\beta'^2\cos^{-2}x_i)\,,
\end{equation}
and thus the spin and the dynamical degrees of freedom decouple.
It is important to note that there is unique minimum $\bx^0=(x_1^0,\dots,x_N^0)$
of the potential $U$ in the Weyl chamber $\tC$, as proved in Ref.~\cite{EFGR04}.

Let
\begin{equation}
\label{Hscalar}
\begin{aligned}
\Hs=-\sum_i \pa_{x_i}^2 &+ a(a-1)\,\sum_{i\neq j}\left(\sin^{-2}
(x_i-x_j)+
\sin^{-2} (x_i+x_j)\right)\\
&{}+b(b-1)\,\sum_i \sin^{-2}\!x_i+b'(b'-1)\,\sum_i
\cos^{-2}\!x_i
\end{aligned}
\end{equation}
denote the Hamiltonian of the scalar $BC_N$ Sutherland model.
The Hamiltonian $\ssh_{\ep}$ of the HS spin chain associated with
$H_\ep$ is defined by
\begin{equation}\label{defssh}
\ssh_\ep=\frac1a\,\big(H_\ep-\Hs\big)\big|_{\bx\,\mapsto\,\bx^0}\,,
\end{equation}
namely
\begin{multline}\label{ssh}
\ssh_{\ep}=\sum_{i\neq j}\Big[\sin^{-2}(x^0_i-x^0_j)\,(1+ S_{ij})
+\sin^{-2}(x^0_i+x^0_j)\,(1+\tS_{ij})\Big]\\
+\sum_i\big(\beta\,\sin^{-2}x^0_i+\beta'\,\cos^{-2}x^0_i\big)(1-\ep
S_i)+8B\sum_i S_i\,.
\end{multline}
This Hamiltonian differs from the one in \cite{EFGR04} by the last term, which
represents the interaction of the spins with a magnetic field along the $\xi$
axis of constant magnitude $-16B/(gq)$.  As shown in Ref.~\cite{EFGR04},
Eq.~\eqref{defssh} and the previous considerations lead to the relation
\begin{equation}\label{ssZ} \ssZ_{\ep}
(T)=\lim_{a\to\infty}\frac{Z_{\ep}(aT)}{\Zs(aT)}\,.
\end{equation}
between the partition functions $\ssZ_\ep$, $Z_\ep$, and $\Zs$
of the respective Hamiltonians  $\ssh_\ep$, $H_\ep$, and $\Hs$.

We shall now compute the partition function of the spin
chain \eqref{ssh} in closed form by evaluating the RHS
of Eq.~\eqref{ssZ}. {}From Eqs.~\eqref{En0} and \eqref{Ens} it follows that
\begin{equation}\label{E}
E_{n\si}\simeq
a^2E_0+8a\Big[\sum_in_i\big(\Bbe+N-i\big)+B\,\la(n,\si)\Big]\,,
\end{equation}
where $\Bbe\equiv\frac12(\beta+\beta')$, and we have dropped
the term independent of $a$ which becomes negligible in the limit $a\to\infty$.
The constant $E_0\equiv4\sum_i(\Bbe+N-i)^2$, which is independent of $n$ and
$\si$, can also be dropped from both partition functions $Z_\ep$ and
$\Zs$ without modifying the value of $\ssZ_\ep$. With this
convention, it was proved in \cite{EFGR04} that when $a\to\infty$ the partition
function of the scalar Sutherland model \eqref{Hscalar} can be
written as
\begin{equation}
\label{Zs} \Zs(aT)\simeq
\prod_i\bigg[1-q^{i\big(\Bbe+N-\tfrac12(i+1)\big)}\bigg]^{-1}\,,
\end{equation}
where we have set
\[
q\equiv\e^{-8/k_\text BT}\,.
\]

The calculation of $Z_\ep(aT)$ is more involved. To perform it, it
is convenient to represent the multiindex $n\in\NN_0^N$ appearing
in \eqref{E} as
\begin{equation} \label{nm}
n =\big(\overbrace{\vphantom{1}m_1,\dots,m_1}^{k_1},
\overbrace{\vphantom{1}m_2,\dots,m_2}^{k_2},\dots,
\overbrace{\vphantom{1}m_r,\dots,m_r}^{k_r}\big)\,,
\end{equation}
where $m_1>\cdots>m_r\geq 0$ and $k_i=\#(m_i)$ satisfies
$\sum_{i=1}^rk_i=N$. The characterization of $\cB_\ep^x$ in the previous section
implies that $k_i\in\{1,2\}$ and $k_r=1$ if $m_r=0$. We shall denote by
$\cP_N$ the set of partitions $k=(k_1,\dots,k_r)$ of the integer
$N$ such that $k_i\in\{1,2\}$. If $k=(k_1,\dots,k_r)\in\cP_N$, we
define
\[
d(k)=\card\big\{i\;\big|\; k_i=1\big\}\,.
\]
Since $d(k)=d_+(n,\si)+d_-(n,\si)$, it follows that
\begin{equation}
\la(n,\si)=2\,d_+(n,\si)-d(k)\,,
\end{equation}
and therefore
\begin{equation}
E_{n\si}\simeq 8a\Big[\ms\sum_i n_i(\Bbe+N-i)+B\big(2d_+(n,\si)-d(k)\big)\Big]\,,
\end{equation}
where we have used Eq.~\eqref{E} without the inessential ground state energy
$E_0$.  After expressing the first sum in terms of $m$ and $k$ as in
Ref.~\cite{EFGR04} we obtain the expression
\begin{equation}\label{E>}
E\simeq 8a\bigg[\ms\sum_{i=1}^rm_i\nu_i(k)+B\big(2d_+(n,\si)-d(k)\big)\bigg]\,,
\end{equation}
where
\[
\nu_i(k)=
k_i\bigg(\Bbe+N-\frac{k_i+1}2-\sum_{j=1}^{i-1}k_j\bigg)\,.
\]
The partition function $Z_\ep(aT)$ is therefore given by
\begin{equation}\label{Zepdef}
Z_\ep(aT)\simeq\sum_{k\in\cP_N}\sum_{m_1>\cdots>m_r\geq 0}
q^{\sum\limits_{i=1}^rm_i\nu_i(k)}\sum_{\ket{\si}} q^{B\ms[2d_+(n,\si)-d(k)]}\,,
\end{equation}
where the sum over the spins is restricted to those values of $\ket\si$ such
that $\La(f_n\ket\si)\in\cB^\xi_\ep$. The latter sum depends essentially on
whether $m_r>0$ or $m_r=0$. Indeed:

\smallskip\noindent\textit{Case 1:} $m_r>0$.

In this case, for a given partition $k\in\cP_N$,
$d_+(n,\si)$ can take any value in the range $0,\dots,d(k)$.
The number of states of the basis $\cB^\xi_\ep$
for which $d_+(n,\si)=\de$ is given by the combinatorial
number $\binom{d(k)}\de$. Hence
\begin{align}
\sum_{\ket{\si}} q^{B\ms[2d_+(n,\si)-d(k)]}&=
q^{-B\ms d(k)}\sum_{\de=0}^{d(k)}\binom{d(k)}\de q^{2B\ms\de}
=q^{-B\ms d(k)}\Big(1+q^{2B}\Big)^{d(k)}\notag\\
&=\Big(q^{B}+q^{-B}\Big)^{d(k)}\,,\qquad (m_r>0)\,.\label{mr>0}
\end{align}

\noindent\textit{Case 2:} $m_r=0$.

Note, first of all, that Condition 1 on the basis $\cB^\xi_\ep$ implies that in this case $k_r=1$.
Let us suppose, to begin with, that $\ep=1$. By condition $3'$ on the basis $\cB^\xi_\ep$
the $\xi$ component of the spin of the last particle must be $\si_N=+1/2$. Thus
$d_+(n,\si)$ must be at least $1$ in this case. Since the value of $\si_N$ is fixed, the
number of states of the basis $\cB^\xi_\ep$
for which $d_+(n,\si)=\de$ is now given by $\binom{d(k)-1}{\de-1}$.
Therefore
\begin{align}
\sum_{\ket{\si}} q^{B\ms[2d_+(n,\si)-d(k)]}&=
q^{-B\ms d(k)}\sum_{\de=1}^{d(k)}\binom{d(k)-1}{\de-1} q^{2B\ms\de}\notag\\
&=q^{B}\Big(q^{B}+q^{-B}\Big)^{d(k)-1}\,,\qquad (m_r=0,\:\ep=1)\,.\label{mr0ep1}
\end{align}
If, on the other hand, $\ep=-1$, the only difference with the case $\ep=1$
is that now the $\xi$ component of the last particle's spin is $\si_N=-1/2$.
Obviously, the value of the sum over the spins can be obtained from Eq.~\eqref{mr0ep1}
by changing the sign of $B$. We thus have
\begin{equation}
\sum_{\ket{\si}} q^{B\ms[2d_+(n,\si)-d(k)]}=
q^{\ep B}\Big(q^{B}+q^{-B}\Big)^{d(k)-1}\,,\qquad (m_r=0)\,.\label{mr0}
\end{equation}
Inserting Eqs.~\eqref{mr>0} and~\eqref{mr0} into the formula~\eqref{Zepdef}
for the partition function we obtain
\begin{multline}\label{Zep}
Z_\ep(aT)\simeq\sum_{k\in\cP_N}
\bigg[\Big(q^{B}+q^{-B}\Big)^{d(k)}\sum_{m_1>\cdots>m_r>0}
q^{\sum\limits_{i=1}^rm_i\nu_i(k)}\\
+\de_{k_r,1}q^{\ep B}\Big(q^{B}+q^{-B}\Big)^{d(k)-1}\sum_{m_1>\cdots>m_{r-1}>0}
q^{\sum\limits_{i=1}^{r-1}m_i\nu_i(k)}\bigg]\,.
\end{multline}
It was shown in Ref.~\cite{EFGR04} that
\begin{equation}\label{sums}
\sum_{m_1>\cdots>m_s>0}q^{\sum\limits_{i=1}^sm_i\nu_i(k)}=\prod_{j=1}^s\frac{q^{N_j}}{1-q^{N_j}}\,,
\end{equation}
with
\[
N_j=\sum_{i=1}^j\nu_i=\bigg(\sum_{i=1}^jk_i\bigg)
\bigg(\Bbe+N-\frac12-\frac12\sum_{i=1}^jk_i\bigg)\,.
\] {}From Eqs.~\eqref{Zep} and~\eqref{sums} it follows that the partition
function $Z_\ep$ of the Hamiltonian~\eqref{H2} satisfies
\begin{equation}\label{Zepfinal}
Z_\ep(aT)\simeq\!\sum_{(k_1,\dots,k_r)\in\cP_N}\!
\Big(q^{B}+q^{-B}\Big)^{d(k)}
\bigg(\prod_{j=1}^{r-1}\frac{q^{N_j}}{1-q^{N_j}}\bigg)
\bigg(\frac{q^{N_r}}{1-q^{N_r}}
+\delta_{k_r,1}\frac{q^{\ep B}}{q^{B}+q^{-B}}\bigg).
\end{equation}
Substituting \eqref{Zs} and \eqref{Zepfinal} into \eqref{ssZ} we finally obtain the
following expression for the partition function of the
Haldane--Shastry spin chain \eqref{ssh}:
\begin{multline}\label{ssZfinal}
\ssZ_\ep(T)=\prod_i\bigg[1-q^{i\big(\Bbe+N-\tfrac12(i+1)\big)}\bigg]
\sum_{(k_1,\dots,k_r)\in\cP_N}\!
\Big(q^{B}+q^{-B}\Big)^{d(k)}
\bigg(\prod_{j=1}^{r-1}\frac{q^{N_j}}{1-q^{N_j}}\bigg)\\
\times\bigg(\frac{q^{N_r}}{1-q^{N_r}}
+\delta_{k_r,1}\frac{q^{\ep B}}{q^{B}+q^{-B}}\bigg).
\end{multline}
It can be seen that all the denominators $1-q^{N_j}$ appearing in this formula
are included as factors in the first product.  Similarly, the denominator in
the last term is always canceled by the factor $\Big(q^{B }+q^{-B}\Big)^{d(k)}$ (indeed, $d(k)\geq1$ for $k_r=1$).  Therefore, as should be
the case for a finite system, the partition function $\ssZ_\ep$ can be written
as a finite sum of terms of the form $d_\mathsf e\,q^{\mathsf e}$, where
$8\mathsf e$ is an eigenvalue of the spin chain Hamiltonian $\ssh_\ep$ and
$d_{\mathsf e}$ its corresponding degeneracy.

The formula~\eqref{ssZfinal} for the partition function $\ssZ_\ep$ reduces to
that found in Ref.~\cite{EFGR04} in the absence of magnetic field.  We also
note that the spin chain Hamiltonians $\ssh_+$ and $\ssh_-$ are no longer
isospectral when $B\neq 0$, although their spectra are obviously related by the
mapping $B\mapsto -B$.


\section*{Acknowledgments}
We would like to dedicate this article to Professor Francesco Calogero in his
70th anniversary. This work was partially supported by the DGI under grant
no.~BFM2002--02646.  A.E. acknowledges the financial support of the Spanish
MEC through an FPU scholarship.

\label{lastpage}
\end{document}